\documentclass[10pt,aps,pra,amssymb,superscriptaddress,onecolumn,manuscript,footinbib]{revtex4-1}
\usepackage{graphicx}
\usepackage{amsmath}
\usepackage{amssymb}
\usepackage{color}
\usepackage{soul}
\usepackage{ulem} 
\usepackage{dsfont}
\usepackage{tikz}
\usepackage{hyperref}

\newcommand{\ket}[1]{|#1\rangle}

\newcommand{\op}[1]{\hat{#1}} 

\newcommand{\abs}[1]{\left|#1\right|}

\begin{document}

\title{Quantum Control in Qutrit Systems using Hybrid Rabi-STIRAP Pulses}

\author{Antti Veps\"al\"ainen}
\affiliation{Low Temperature Laboratory, Department of Applied Physics, Aalto University School of Science, P.O. Box 15100, FI-00076 AALTO, Finland}

\author{Sergey Danilin}
\affiliation{Low Temperature Laboratory, Department of Applied Physics, Aalto University School of Science, P.O. Box 15100, FI-00076 AALTO, Finland}

\author{Elisabetta Paladino}
\affiliation{Dipartimento di Fisica e Astronomia, Universit\`a di Catania, Via
Santa Sofia 64, 95123 Catania, Italy}
\affiliation{CNR-IMM UOS Universit\`a (MATIS), Consiglio Nazionale delle
Ricerche, Via Santa Sofia 64, 95123 Catania, Italy, \& INFN,
Sezione di Catania, Via Santa Sofia 64, 95123 Catania, Italy}

\author{Giuseppe Falci}
\affiliation{Dipartimento di Fisica e Astronomia, Universit\`a di Catania, Via
Santa Sofia 64, 95123 Catania, Italy}
\affiliation{CNR-IMM UOS Universit\`a (MATIS), Consiglio Nazionale delle
Ricerche, Via Santa Sofia 64, 95123 Catania, Italy, \& INFN,
Sezione di Catania, Via Santa Sofia 64, 95123 Catania, Italy}

\author{Gheorghe Sorin Paraoanu}
\affiliation{Low Temperature Laboratory, Department of Applied Physics, Aalto University School of Science, P.O. Box 15100, FI-00076 AALTO, Finland}

\begin{abstract}We introduce and analyze theoretically a procedure that combines slow adiabatic STIRAP manipulation with short nonadiabatic Rabi pulses to produce any desired three-level state in a qutrit system. In this protocol, the fast pulses create superpositions between the ground state and the first excited state, while the slow pulses transfer an arbitrary population to the second excited state via stimulated Raman adiabatic passage (STIRAP). We demonstrate high-fidelity quantum control of the level populations and phases and we characterize the errors incurred under the breakdown of adiabaticity. In a configuration where an ancillary state is available, we show how to realize a nondemolition monitoring of the relative phases. These methods are general and can be implemented on any experimental platform where a quantum system with at least three accessible energy levels is available. We discuss here in detail experimental implementations in circuit quantum electrodynamics (QED) based on the results obtained with a transmon, 
where the control of population using the hybrid Rabi-STIRAP sequence has been achieved.
\end{abstract}

\maketitle

\section{Introduction}
 Using multi-level quantum systems instead of the commonly employed two level qubits extends the Hilbert space of the system, which in turn would reduce the number of elements needed to perform a given computational task \cite{Toffoli,Lanyon2009,qutrit_logic_luo}. However, this advantage comes at the cost of increased requirements for the accuracy of the control pulses used to manipulate the states of the system. A key operation in quantum computation is the efficient and robust preparation of the initial state, which serves as a starting point of quantum algorithms. For example, in a two-level system the application of a $\pi$-pulse switches the qubit from the ground state to the first excited state, and any superposition can be easily created by applying Rabi pulses with appropriate length and amplitude. The task becomes more complicated if we intend to control a system with three states $|0\rangle$, $|1\rangle$, and $|2\rangle$ because some of the transitions might be forbidden due to selection rules. In this case, one has to transfer the population in some other way. A simple approach would be to apply a sequence of pulses, first a $\pi_{01}$ pulse followed by a $\pi_{12}$ pulse: however, this sequence will be quite sensitive to the timing and the amplitude of the pulses. This makes the generation of such a sequence an experimentally demanding task -  in general susceptible to environmental fluctuations and instrumentation errors.

However, the state preparation can be made robust by using adiabatic control pulse sequence. The adiabatic population transfer is based on modifying the parameters of the eigenstates of the system. If the change in the eigenstates of the systems is slow enough, the system remains in the instantaneous ground state of the Hamiltonian, which is adiabatically (transition-free) modified by the change of the parameters. The stimulated Raman adiabatic passage (STIRAP) algorithm \cite{reviewSTIRAP,another_reviewSTIRAP} realizes the population transfer by employing two slowly varying, temporally slightly overlapping control pulses applied on 12 and 01 transitions. STIRAP acts only in the subspace $\{|0\rangle,|2\rangle\}$ of the initial state $|0\rangle$ and target state $|2\rangle $, and it does not populate the intermediate state $|1\rangle$. This can be understood as a destructive interference on the state $|1\rangle$, together with the formation of a zero-eigenenergy dark state in the subspace $\{|0\rangle,|2\rangle\}$. In a full STIRAP sequence, the system follows adiabatically the dark state, starting from the ground state and ending in the second excited state.

In recent years the field of superconducting qubits has experienced a fast experimental progress, and state-preparation techniques such as those mentioned above became more and more relevant for the aim of realizing high-fidelity state preparation in quantum processors. Various experiments demonstrated already that three-level and multilevel systems can be realized using superconducting circuits based on the Josephson effect \cite{AT_wallraff,coherent_population_trapping,EIT_abdumalikov,AT_us,ourPRB,dynamicalAT,ka:215-peterergustaffson-prl-transmohhigherlevel,fedorov}. Several theoretical proposals addressed the implementation of STIRAP in these devices~\cite{kr:211-younori-nature-multilevel,stirap_Paladino}, including Cooper
pair boxes~\cite{pino_PRB2009,lambda_quantronium_falci,ka:216-distefano-pra-twoplusone}. In the case of a transmon \cite{transmon_PRA2007}, STIRAP becomes relevant because  the direct transfer of population from the ground state to the second excited state is forbidden in the first order due to a vanishingly small electric dipole moment between these two states. Here we demonstrate a method to fully control the complex coefficients of the wave-function, by using a combination of Rabi pulses and STIRAP.
We show that if the STIRAP sequence is preceded by a non-adiabatic pulse on the 0 - 1 transition, it is possible to create an arbitrary initial state in the full space spanned by $\{|0\rangle,|1\rangle,|2\rangle\}$ states. Earlier this method has been studied in \cite{stirap_ours}, but only the control of the absolute values of the wave-function amplitudes (state populations) has been demonstrated. The result is important for the field of quantum control and in particular for analog quantum simulation \cite{ParaoanuJLTP}, opening an alternative route to the emulation of large-spin systems \cite{Martinis2009}. Our method can be extended to multilevel systems by combining Rabi pulses with multiple adiabatic pulses ({\it e.g.} straddle STIRAP) \cite{PhysRevA.56.4929,Unanyan1998144,PhysRevA.63.043401,0953-4075-36-5-310,Vitanov200155}.

The paper is organized as follows: in Section 2 we present our main results regarding a qutrit under a hybrid Rabi-STIRAP pulse. We derive an analytical expression for the wavefunction at an arbitrary time and we put in evidence the role of the phases. We study the populations and the relative phases between the three states as a function of the length of the Rabi pulse and of the width of the STIRAP pulses. We show that a robust, nonoscillating  behaviour is obtained only under the condition of adiabaticity for STIRAP. In Section 3 we analyze the case of an additional fourth state, and demonstrate a protocol where this state is used to monitor the phases between the initial state and the target state. In Section 4 we present a more in-depth experimental evaluation of the transmon for implementing the proposed experiments. Finally, our conclusions are presented in Section 5 and a further generalization of the results of Section 2 is delegated to Appendix A.

\section{Hybrid quantum control in qutrits}

It is always possible to combine diabatic Rabi pulses and STIRAP sequence: the question is if the final effect of the combined pulse can be understood in a simple enough way, such that, by tuning one parameter of either the Rabi or STIRAP pulse we are able to give a simple recipe - easy to implement experimentally -  for producing the desired tripartite state. It is not obvious that this is possible: indeed, the first Rabi pulse, acting on the $0-1$ transition, will produce an occupation of the first excited state. Therefore the subsequent STIRAP-like pulse sequence does not stabilize a dark state, the whole process involving both destructive and constructive interference in a non-obvious combination. A related question is the impact on interference  caused by the manipulation of the  phases of the fields, which is a key ingredient for efficient implementation of general rotations in the Hilbert space. In this section we give an analytical treatment of this problem using the adiabatic approximation, supported by numerical results accounting for effects of nonadibaticity.
\begin{figure}[tb]
\centering
\includegraphics[width=15cm]{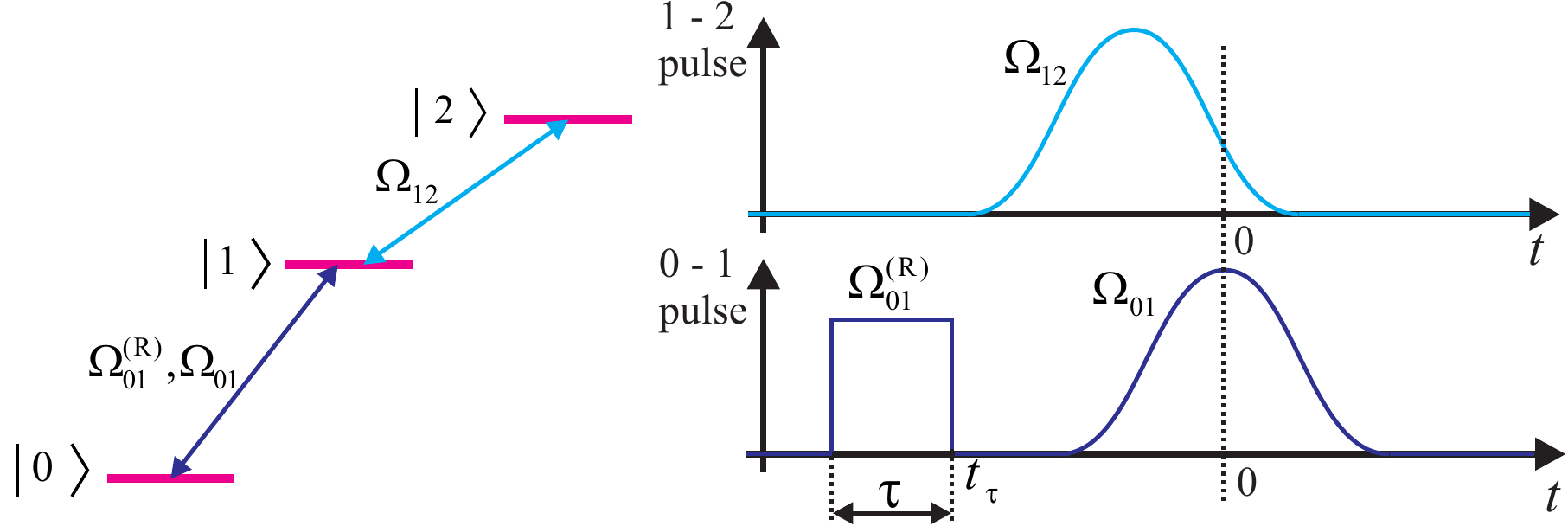}
\caption{Schematic of the hybrid pulse sequence for a three-level system in the ladder configuration and with the two transitions driven resonantly by two fields. First, a superposition between state $|0\rangle$ and state $|1\rangle$ is created by a nonadiabatic Rabi pulse $\Omega_{01}^{(\mathrm{R})}$ (for simplicity taken as a square pulse). Then a STIRAP sequence is applied, with a pulse on the 1-2 transition (with coupling strength $\Omega_{12}$) followed by another pulse on the 0-1 transition (with coupling strength $\Omega_{01}$).
}
\label{fig:hybrid_2d.jpg}
\end{figure}

Our Hamiltonian takes the standard form for a driven three-level system in the rotating-wave approximation and with the two drives on-resonance with the corresponding transitions
\begin{equation}
\label{eq:threelevel}
\op{H} = \frac{\hbar}{2}
\begin{bmatrix}
0 & |\Omega_{01}|e^{i \phi_{01}} & 0 \\
|\Omega_{01}|e^{- i \phi_{01}} & 0 &
|\Omega_{12}| e^{i \phi_{12}}\\
0 & |\Omega_{12}| e^{-i \phi_{12}}& 0
\end{bmatrix}.
\end{equation}
The eigenvalues of this Hamiltonian are $\lambda_\pm = \pm \frac{\hbar}{2} \Omega$, where $\Omega = \sqrt{|\Omega_{01}|^2 + |\Omega_{12}|^2}$ and $\lambda_{0} = 0$, corresponding to eigenvectors
\begin{eqnarray}
\label{eq:awg_states}
\ket{n_{\pm}} &=&\frac{1}{\sqrt{2}}\ket{\mathcal{B}} \pm \frac{e^{-i\phi_{01}}}{\sqrt{2}}\ket{1}, \\
\ket{n_{0}} &=& \ket{\rm {\cal D}},
\end{eqnarray}
where the dark and bright states are defined as two orthogonal states,
\begin{eqnarray}
\label{eq:darkbright}
\ket{\mathcal{B}} &=& \sin\Theta \ket{0} +  e^{-i\phi_{01}-i\phi_{12}}\cos\Theta\ket{2},\\
\ket{\mathcal{\cal D}} &=& \cos\Theta \ket{0} -   e^{-i\phi_{01}-i\phi_{12}}\sin\Theta\ket{2}.
\end{eqnarray}
Note that in the rotating-wave approximation the Hamiltonian in Eq. (\ref{eq:threelevel}) is time-dependent.
Here the STIRAP angle $\Theta$ has the expression  $\tan{\Theta} = |\Omega_{01}|/|\Omega_{12}|$, or in other words we parametrize the couplings by $|\Omega_{01}|=\Omega \sin \Theta$ and  $|\Omega_{12}|=\Omega \cos \Theta$.

The STIRAP sequence starts with an initial state created by the Rabi pulse (see Fig. \ref{fig:hybrid_2d.jpg})
\begin{equation}
|\psi (t_{\tau})\rangle = \alpha |0\rangle + \beta |1\rangle,
\end{equation}
where $\alpha$ and $\beta$ are the complex coefficients prepared by the $0 - 1$ pulse of duration $\tau$ and amplitude $\Omega^{(R)}_{01}$.
This state can be further rewritten in the basis of instantaneous eigenvectors, and for all practical purposes we can say that the STIRAP sequence starts at the time $t_{\tau}$ when the $0 - 1$ pulse ends, from the state
\begin{equation}
|\psi (t_{\tau})\rangle = \alpha |n_{0} (t_{\tau}) \rangle + \frac{\beta}{\sqrt{2}}e^{i \phi_{01}} \left( |n_{+} (t_{\tau})\rangle -
|n_{-} (t_{\tau})\rangle \right).
\end{equation}
In the adiabatic approximation and for times $t > t_{\tau}$ this state evolves as
\begin{equation}
|\psi (t)\rangle = \alpha e^{i \zeta_{0} (t)} |n_{0} (t) \rangle + \frac{\beta}{\sqrt{2}}e^{i \phi_{01}} \left( e^{i \zeta_{+} (t)}|n_{+} (t)\rangle -
e^{i \zeta_{-} (t)}|n_{-} (t)\rangle \right).
\end{equation}
Here the phases $\zeta_{k}$, with $k \in \{0,+,-\}$, comprise a dynamical rotation at the frequencies corresponding to the eigenvalues $\lambda_{k}$ and a geometrical phase $\gamma_{k} (t)$,
\begin{eqnarray}
\zeta_{k} (t) &=& -\frac{1}{\hbar} \int_{t_{\tau}}^{t} dt' \lambda_{k} (t') + \gamma_{k} (t), \\
\gamma_{k}(t) &=& i \int_{t_{\tau}}^{t} dt' \langle n_{k} (t') |\partial_{t' } n_{k} (t') \rangle .
\end{eqnarray}

\begin{figure}[tb]
\centering
\includegraphics[width=\textwidth]{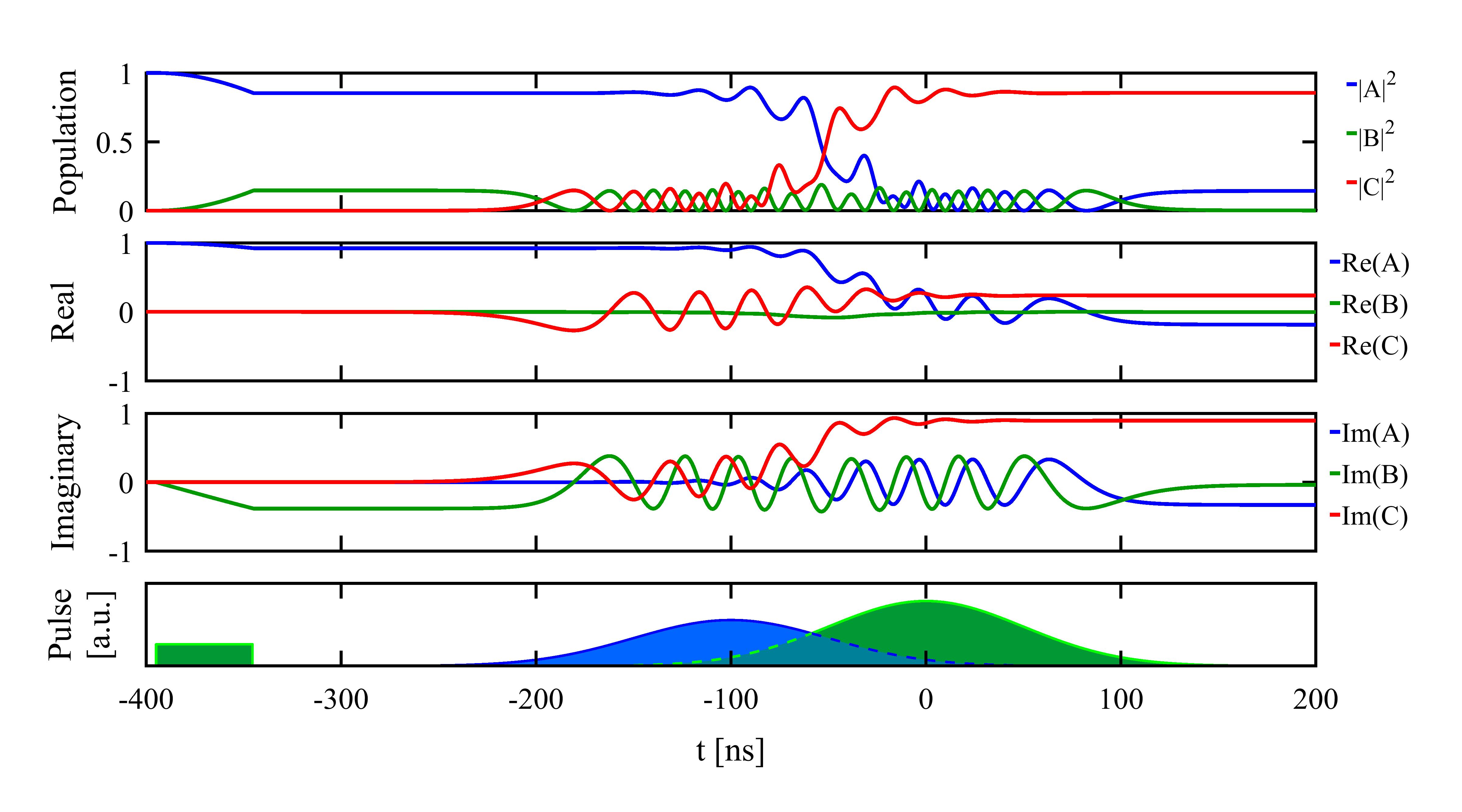}
\caption{Time evolution of the system. The upper panels show the populations of the three states $|A|^2$, $|B|^2$, and $|C|^2$, as well as real and imaginary parts of the the wavefunction amplitudes during the hybrid STIRAP sequence. The pulses driving the sequence are shown in the lowest panel. The simulation is performed with the parameters $\phi_{01} = \pi/3$, $\phi_{12} = \pi/4$, $\Omega_{01}^{(0)} = \Omega_{12}^{(0)} = 37.5$ MHz, $\sigma = 50$ ns, $t_{\rm s}/\sigma = 2$, $\alpha = \cos(\pi/8)$, and $\beta = -i\sin(\pi/8)$.}
\label{fig:time_trajectories}
\end{figure}

If the phases $\phi_{01}$ and $\phi_{12}$ of the STIRAP pulses are time-independent, then  $\langle n (t') |\partial_{t' } n (t') \rangle =0$ for all the $n$'s, and as a result we are left with
\begin{equation}
|\psi (t)\rangle = \alpha |n_{0} (t) \rangle + \frac{1}{\sqrt{2}}\beta e^{-\frac{i}{2} \int_{t_{\tau}}^{t} dt' \Omega(t')+ i \phi_{01}}|n_{+} (t)\rangle -
\frac{1}{\sqrt{2}}\beta e^{+\frac{i}{2} \int_{t_{\tau}}^{t} dt' \Omega(t')+ i \phi_{01}}|n_{-} (t)\rangle .
\end{equation}
In the original basis $\{|0\rangle, |1\rangle, |2\rangle \}$ this reads
\begin{eqnarray}
|\psi (t)\rangle &=& \left[ \alpha \cos \Theta (t) - i \beta e^{i \phi_{01}}\sin \Theta (t) \sin \left(\frac{1}{2} \int_{t_{\tau}}^{t}dt' \Omega (t' ) \right)\right] |0\rangle \notag\\
& & + \beta \cos \left(\frac{1}{2}\int_{t_{\tau}}^{t}dt' \Omega (t' )\right)|1\rangle \notag\\
& &  -\left[ \alpha e^{-i \phi_{01}}\sin \Theta (t)  + i \beta \cos \Theta (t) \sin \left(\frac{1}{2} \int_{t_{\tau}}^{t}dt' \Omega (t' ) \right)\right]e^{-i \phi_{12}} |2\rangle . \label{fracSTIRAP}
\end{eqnarray}
 After the full STIRAP ($t \rightarrow \infty$, or $\Theta = \pi /2$) we obtain

\begin{figure}[tb]
\centering
\includegraphics[width=\textwidth]{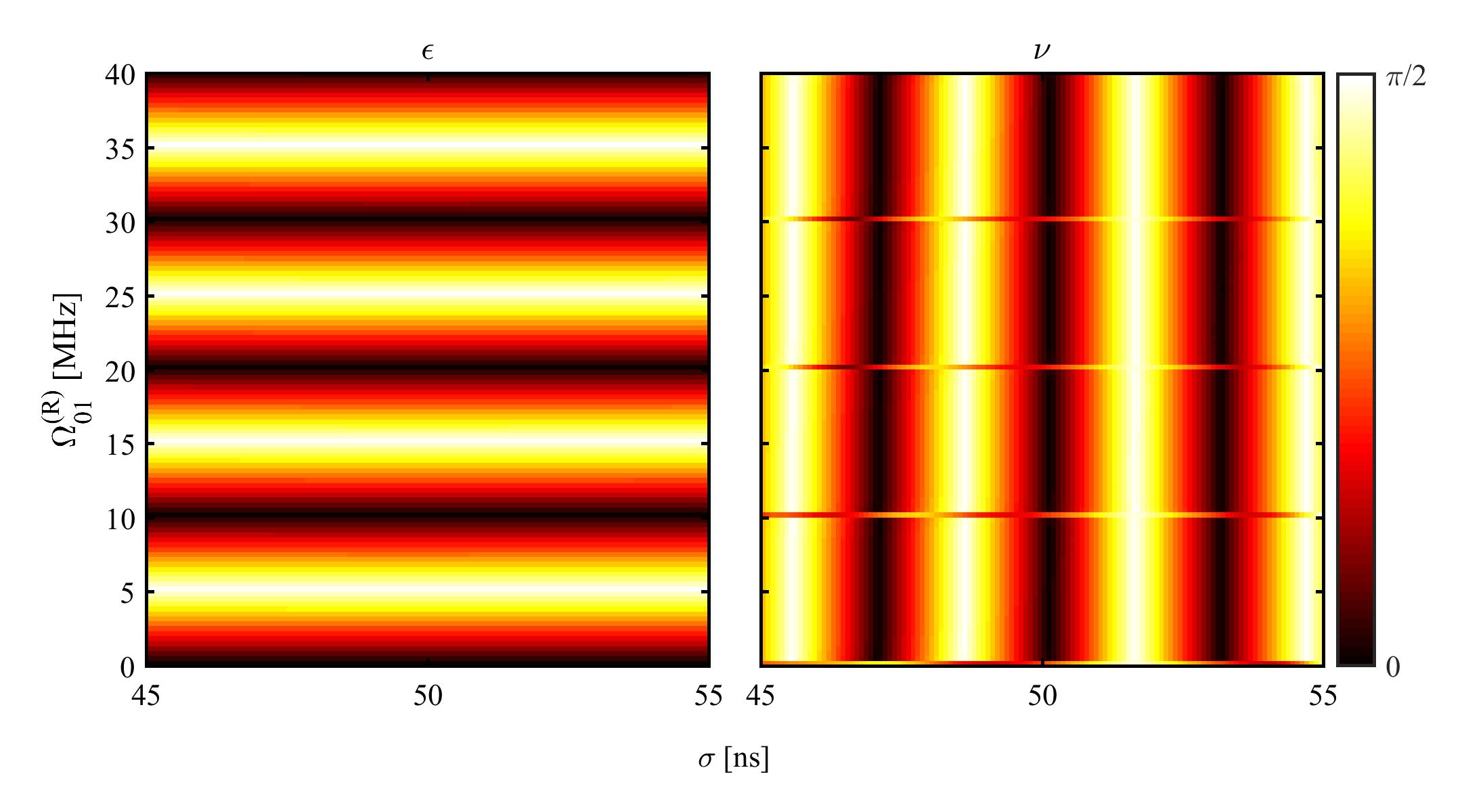}
\caption{The magnitude of the wavefunction coefficients in spherical coordinates, $\epsilon$ and $\nu$, as a function of the Rabi pulse amplitude $\Omega^{(R)}_{01}$ and the duration of the STIRAP pulses $\sigma$. The horizontal lines in the right plot result from the numerical instability when $|A| = |B| = 0$ and $\nu$ is undefined. The parameters of the simulation are $\phi_{01} = \pi/3$, $\phi_{12} = \pi/4$, $\Omega_{01}^{(0)} = \Omega_{12}^{(0)} = 37.5$ MHz, $t_{\rm s}/\sigma = 2$, and $\sigma = 50$ ns. The simulation agrees with the adiabatic-approximation analytical result Eq. (\ref{eq:psi_inf}).
}
\label{fig:Ad01r_vs_sigma_sphr}
\end{figure}

\begin{figure}[tb]
\centering
\includegraphics[width=\textwidth]{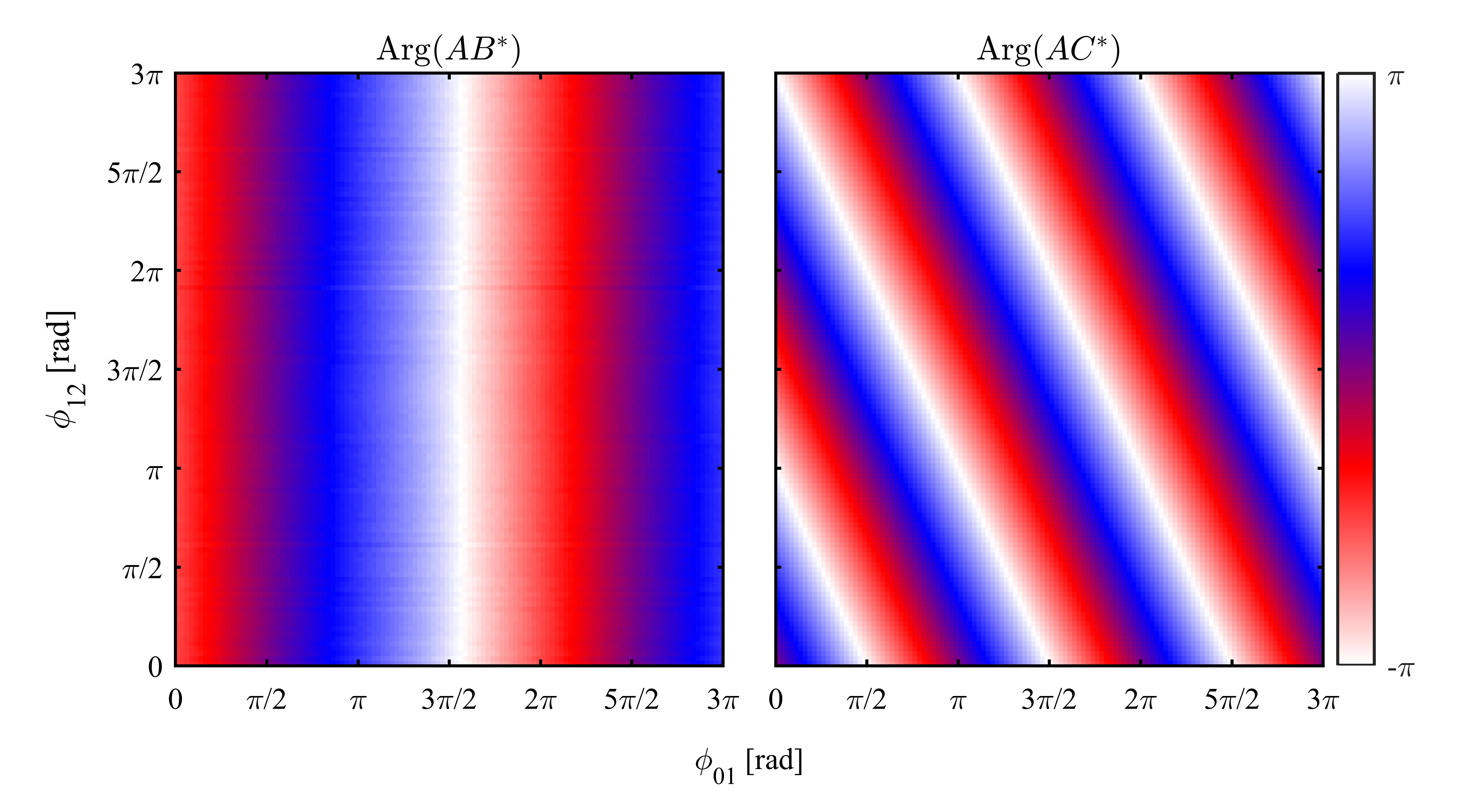}
\caption{The relative phases between the complex amplitudes $A$, $B$, and $C$ of the qutrit wavefuction as a function of the STIRAP phases $\phi_{01}$ and $\phi_{12}$, with $\alpha = \cos(\pi/8)$, $\beta = -i\sin(\pi/8)$, $\Omega_{01}^{(0)} = \Omega_{12}^{(0)} = 37.5$ MHz, $t_{\rm s}/\sigma = 2$, and $\sigma = 50$ ns. The simulation agrees with the analytical results Eqs. (\ref{relphaseAB}, \ref{relphaseAC}) obtained in the adibatic approximation.}
\label{fig:phi01vsphi12}
\end{figure}



\begin{equation}
\label{eq:psi_inf}
|\psi (\infty )\rangle = - i \beta e^{i \phi_{01}}\sin \left(\frac{1}{2} \int_{t_{\tau}}^{\infty}dt' \Omega (t' ) \right) |0\rangle
+  \beta \cos \left(\frac{1}{2}\int_{t_{\tau}}^{\infty}dt' \Omega (t' )\right)|1\rangle - \alpha e^{-i (\phi_{01}+\phi_{12})}|2\rangle .
\end{equation}
This leads to the creation of a state with fully adjustable complex amplitudes, which are controlled by tuning the coefficients $\alpha$ and $\beta$, the phases $\phi_{01}$ and $\phi_{12}$, and the area of the envelopes $\int_{t_\tau}^{\infty} \Omega(t)\mathrm{d}t$. The result shows that the population in state $|2\rangle$ does not depend on the phases or on the STIRAP couplings $\Omega_{01}$ and $\Omega_{12}$. It only depends on the population left on the state $|0\rangle$ immediately after the Rabi preparation pulse is applied to the transition $|0\rangle \rightarrow |1\rangle$. This provides a useful robustness feature, since experimentally it is often the case that the $1-2$ transition is more  difficult to control. In contrast, if one wishes to create general qutrit states by applying sequences of Rabi pulses, the population on the state $|2\rangle$ will accumulate errors from the timing and Rabi frequency of the pulse applied to the $1 -2$ transition. Also the phase differences between any two states can be controlled independently by the STIRAP phases $\phi_{01}$ and $\phi_{12}$, up to a $\pi$-jump resulting from the change of sign of $\sin \left(\frac{1}{2}\int_{t_{\tau}}^{\infty}dt' \Omega (t' )\right)$ and the corresponding cosine terms. In consequence, we can almost independently tune the magnitudes and the phases of the resulting state. The result also can be generalized to an arbitrary initial (at the input of the STIRAP sequence) superposition of the states $|0\rangle$, $|1\rangle$, and $|2\rangle$ (see Appendix).



To investigate this result in more detail, we simulate numerically the time evolution of the system using Eq. \eqref{eq:threelevel} with STIRAP pulses having Gaussian envelopes
\begin{equation}
\begin{aligned}
|\Omega_{01}(t)| &= \Omega_{01}^{(0)}e^{-\frac{t^2}{2\sigma^2}}, \\
|\Omega_{12}(t)| &= \Omega_{12}^{(0)}e^{-\frac{(t + t_s)^2}{2\sigma^2}}, \\
\end{aligned}
\end{equation}
where the dimensionless pulse separation $t_s /\sigma = 2$, $\int_{-\infty}^\infty \Omega^{(R)}_{01}(t) \mathrm{d}t = \pi/4$, and $\int_{-\infty}^\infty |\Omega_{01}(t)| \mathrm{d}t = \int_{-\infty}^\infty |\Omega_{12}(t)| \mathrm{d}t \approx 19\pi$, giving $\int_{t_{\tau}}^{\infty} \Omega (t) \mathrm{d}t= 32.90\pi$. The result is shown in Fig. \ref{fig:time_trajectories}, where we have used the following general expression for the qutrit state
\begin{eqnarray}
|\psi \rangle &=& A|0\rangle + B|1\rangle + C|2\rangle \\
& =&  |A|e^{i\mathrm{Arg}(A)}\ket{0} + |B|e^{i\mathrm{Arg}(B)}\ket{1} + |C|e^{i\mathrm{Arg}(C)}\ket{2}.
\end{eqnarray}
In the ideal case where there are no diabatic losses in the STIRAP process, the simulation replicates the results of Eq. \eqref{fracSTIRAP}. However, in any real process there will exist transitions between the states in the instantaneus basis leading to deviations from the adiabatic time evolution.

In order to create an arbitrary final state, we need to obtain simple relations between the coefficients $A$, $B$, and $C$ and the parameters of the control pulses. Because the absolute values and the arguments of the complex wavefunction coefficients can be independently varied, we can solve the problem in two parts: first we address the absolute values of the coefficients, and in the next step we analyze the phase differences. Since the final state is normalized, we have $|A|^2 + |B|^2 + |C|^2 =1$, and the absolute values of the amplitudes lie on the surface of a sphere. This suggests a parametrization with angles $\epsilon$ and $\nu$ in spherical coordinates,
\begin{eqnarray}
|A| &=& \sin \nu \sin \epsilon ,\\
|B| &=& \cos \nu \sin \epsilon ,\\
|C| &=& \cos \epsilon ,
\end{eqnarray}
or $\epsilon = \arccos(|C|)$ and  $\nu = \arctan\left(|A|/|B|\right)$, where both $\nu,\epsilon \in [0,\pi/2]$. In these coordinates, the spherical angles $\nu$ and $\epsilon$ can be changed
in a simple way, as presented in the simulation of Fig. \ref{fig:Ad01r_vs_sigma_sphr}. We observe that the angle $\epsilon$ does not depend on $\sigma$, thus confirming the analytical result of Eq. (\ref{eq:psi_inf}). On the other hand, the angle $\nu$ depends only on the width $\sigma$ of the STIRAP pulses and does not depend on $\Omega^{(\mathrm{R})}_{01}$.

Next, we can characterize the relative phases of the amplitudes, defined as $\mathrm{Arg}(AB^*)$ and $\mathrm{Arg}(AC^*)$.
As shown in Fig. \ref{fig:phi01vsphi12}, by changing $\phi_{01}$ and $\phi_{12}$ it is possible to achieve all the combinations for the phases. Another option would be to tune the phase of the initial Rabi pulse and $\phi_{01}$ of the STIRAP pulse, whereas fixing $\phi_{01}$ and changing the phase of the Rabi pulse and $\phi_{12}$ is not enough to produce all the required combinations.  The dependence of the relative phases of the amplitudes as obtained from Eq. \eqref{eq:psi_inf} satisfy the simple relations
\begin{eqnarray}
\mathrm{Arg}(AB^*)&=& -\pi/2 + \phi_{01} + \frac{\pi}{2}\left[1-\textrm{sgn} \left[\sin \left(\int_{t_{\tau}}^{\infty}dt' \Omega (t' ) \right)\right]\right] , \label{relphaseAB}\\
\mathrm{Arg}(AC^*) &=& \pi/2 + \mathrm{Arg}(\beta) - \mathrm{Arg}(\alpha)+2\phi_{01} + \phi_{12}+ \frac{\pi}{2}\left[1-\textrm{sgn} \left[\sin \left(\frac{1}{2} \int_{t_{\tau}}^{\infty}dt' \Omega (t' ) \right)\right]\right].\label{relphaseAC}
\end{eqnarray}

\begin{figure}[tb]
\centering
\includegraphics[width=15cm]{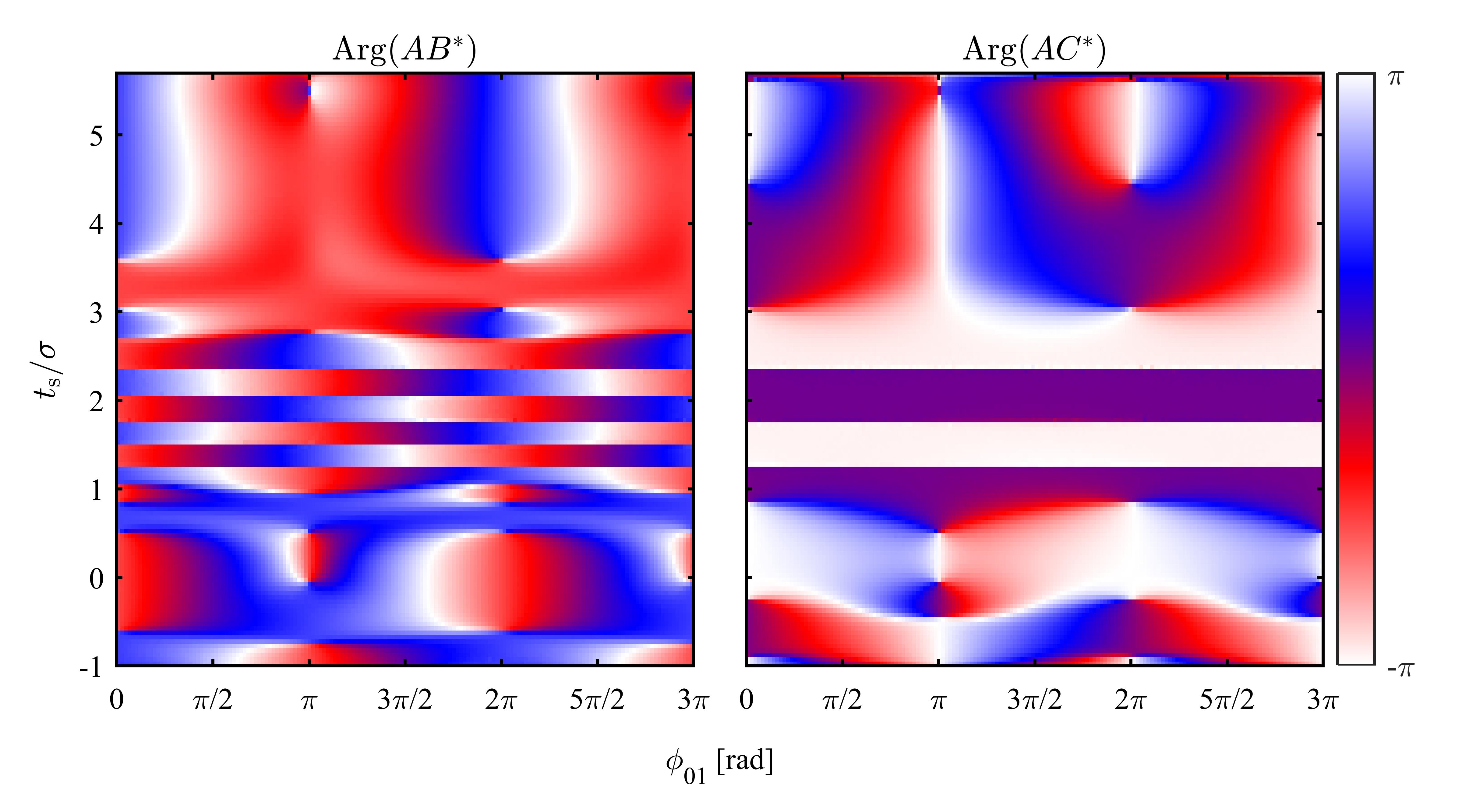}
\caption{The relative phases between the complex amplitudes $A$, $B$, and $C$ of the qutrit as a function of the dimensionless STIRAP pulse separation $t_s/\sigma$ and the STIRAP phases $\phi_{01}$ and $\phi_{12}$. The STIRAP  phases are changed so that $2\phi_{01} + \phi_{12} = 4\pi$. When the adiabatic condition $t_{\rm s}/\sigma \in [1,3]$ is met, the relative phases are given by Eqs. (\ref{relphaseAB}, \ref{relphaseAC}), resulting in the formation of a stripe structure. The abrupt $\pi$ phase shifts across the stripes are due to the changes of sign of the amplitudes, as can be seen from
the $\sin$ and $\cos$ terms of Eq. (\ref{eq:psi_inf}). The parameters for the simulation are $\Omega_{01}^{(0)} = \Omega_{12}^{(0)} = 37.5$ MHz, $\sigma = 50$ ns, $\alpha = \cos(\pi/8)$, and $\beta = -i\sin(\pi/8)$.}
\label{fig:phi01vsk}
\end{figure}

The above rules for creating the desired state apply as long as the adiabatic condition is satisfied. When the adiabatic condition breaks, the situation changes as is demonstrated in Fig. \ref{fig:phi01vsk}. The adiabaticity of the population transfer depends on the overlap of the driving pulses \cite{stirap_overlap}. By changing the dimensionless pulse separation $t_{\rm s}/\sigma$ we can move from adiabatic time evolution to non-adiabatic regime, and observe whether the relations Eqs. (\ref{relphaseAB}, \ref{relphaseAC}) are satisfied.
The adiabatic condition is approximately met while $t_{\rm s}/\sigma \in [1,3]$. In these regions one notices the formation of stripe structures, with the dependence on $\phi_{01}$ on each stripe similar to that presented in Fig. 3 for $t_{\rm s}/\sigma =2$. Along the $t_{\rm s}/\sigma$ direction, the neighboring stripes differ from each other by a factor of $\pi$. This is due to the fact that the qutrit waveform amplitudes change sign, as captured by the last two terms in Eqs. (\ref{relphaseAB}, \ref{relphaseAC}). Note that, as we change $t_{\rm s}/\sigma$, the number of stripes in $\mathrm{Arg}(AB^*)$ is twice as large as the number of stripes in $\mathrm{Arg}(AC^*)$, again in agreement with Eqs. (\ref{relphaseAB}, \ref{relphaseAC}).
The breaking of the phase relation outside $t_{\rm s}/\sigma \in [1,3]$ shows that the adiabatic result of Eq. \eqref{eq:psi_inf} is no longer valid.

\section{Hybrid pulse sequence for a three-level system with additional ancillary state}

\begin{figure}[tb]
\centering
\includegraphics[width=15cm]{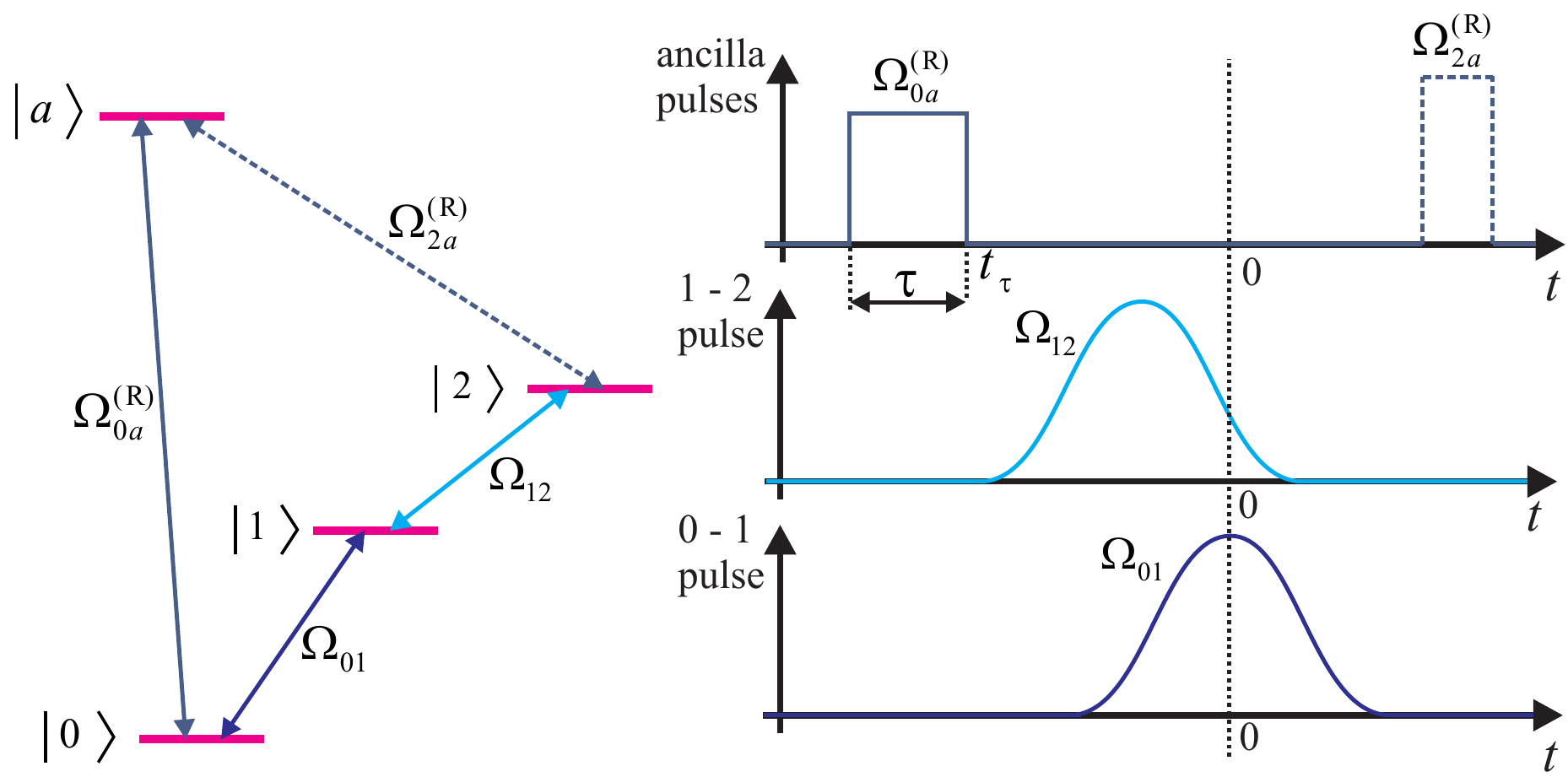}
\caption{Schematic of the three-level system augmented by a fourth ancillary state $|a\rangle$. The nonadiabatic pulse couples the states $|0\rangle$ and $|a\rangle$ and the two STIRAP pulses drive the transitions  0-1 and 1-2.}
\label{schematic_hybridSTIRAP_ancillary}
\end{figure}

We discuss here the case when an auxiliary level $|a\rangle$ exists, see Fig. \ref{schematic_hybridSTIRAP_ancillary}. For example, one can use the fourth level of the artificial atom or the Jaynes-Cummings ladder if control at the single-photon level is achieved in the system. In this case the Hilbert space is large enough to separate the  nonadiabatic Rabi pulse and the STIRAP.

The state at a time $t$ reads in this case
\begin{equation}
|\psi (t)\rangle = \alpha \cos \Theta (t)|0\rangle - \alpha \sin \Theta (t) e^{-i (\phi_{01}+ \phi_{12})} |2\rangle
+ \beta |\mathrm{a}\rangle .  \label{eq:a_final}
\end{equation}
In this configuration, STIRAP and the initial Rabi pulse on the $0 - a$ transition with Rabi coupling $\Omega_{0a}^{(\rm R)}$ allow us to create an arbitrary superposition in the subspace $\{ |0\rangle , |2\rangle , |a\rangle \}$. The populations in the three states $|0\rangle$ , $|2\rangle$ , and $|a\rangle$ are given by $|\alpha|^2 \cos^2 \Theta$, $|\alpha|^2 \sin^2 \Theta$, and $|\beta |^2$ respectively.  The relative phases of the amplitudes are controlled by the phases of $\alpha$ and $\beta$ together with the STIRAP phases.

Besides the realization of superpositions in the $\{ |0\rangle , |2\rangle , |a\rangle \}$ subspace, this configuration can be used for the nondestructive monitoring of the STIRAP phases, as detailed below. Suppose that we interrupt the STIRAP sequence (see Fig. \ref{schematic_hybridSTIRAP_ancillary}) at some mixing angle $\Theta$, and at the end of the sequence depicted in Fig. \ref{schematic_hybridSTIRAP_ancillary} we apply another Rabi pulse $\Omega_{2a}^{(\rm R)}$ that couples levels $|2\rangle$ and $|a\rangle$. In general, an unitary acting only on $\{ |2\rangle , |a\rangle \}$ has the form
\begin{equation}
U' = |0\rangle \langle 0| + \alpha ' |a\rangle \langle a| + \beta ' |a\rangle \langle 2| - e^{-i\phi '}\beta ' |2\rangle \langle a| + e^{i \phi '} \alpha '^{*} |2\rangle \langle 2|,
\end{equation}
with normalization $|\alpha '|^{2} + |\beta '|^{2} = 1$.
Applying this operator to Eq. (\ref{eq:a_final}) we obtain
\begin{eqnarray}
U' |\psi \rangle &=& \alpha \cos \theta |0\rangle - \left[\alpha \alpha '^{*} \sin\Theta e^{-i (\phi_{01} + \phi_{12} - \phi')} + \beta \beta 'e^{-i \phi '}\right] |2\rangle \\
& & - \left[\alpha\beta ' \sin \Theta e^{-i (\phi_{01} + \phi_{12})} - \beta \alpha '\right] |a\rangle .
\end{eqnarray}
One sees the formation of an interference structure in the amplitude corresponding to the ancillary state $|a\rangle$. Let us consider from now on $\phi' =0$ and real $\alpha = |\alpha |$, $\alpha' = |\alpha'|$, $\beta = |\beta |$, and $\beta' = |\beta '|$. Then the occupation probability
of the ancillary state is $P_{a}=\alpha ^2 \beta'^{2} \sin^{2}\Theta + \beta^{2}\alpha'^{2} - 2 \alpha \beta \alpha'\beta'\sin \Theta \cos (\phi_{01} + \phi_{12} )$. Measuring the population of the ancillary state would then reveal an interference pattern as a function of $\phi_{01} + \phi_{12}$ with visibility
\begin{equation}
v = \frac{P_{a}^{(\rm max )} - P_{a}^{(\rm min )}}{P_{a}^{(\rm max )} + P_{a}^{(\rm min )}} = \frac{2 \alpha \beta \alpha ' \beta ' \sin \Theta }{\alpha ^{2} \beta'^{2}\sin^{2} \Theta + \beta ^{2} \alpha'^{2}} .
\end{equation}
Let us consider now the case $\beta \ll 1$ and $\beta ' \ll 1$. We obtain $U' |\psi \rangle \approx \cos \Theta |0\rangle - \sin\Theta e^{-i (\phi_{01} + \phi_{12})} |2\rangle$, in other words the STIRAP is unaffected by pulses that address the ancilla state. However, the visibility remains large,
\begin{equation}
v \approx \frac{2\beta \beta ' \sin \Theta }{\beta'^{2}\sin^{2} \Theta + \beta ^{2}} ,
\end{equation}
and can reach even the value $v = 1$ if we arrange the Rabi pulses such that $\beta = \beta' \sin \Theta$. This technique can be immediately extended to the detection of phases produced by Hamiltonians more general than Eq. (\ref{eq:threelevel}); for example it can be used for the detection of Berry phases and of the phases involved in holonomic quantum gates \cite{geometricphasegate,berrywallraff,nonabelianwallraff}.

\section{Experimental implementation in circuit QED}

The protocols presented here can be implemented in any quantum system that has three or more accessible and coherent energy levels (atoms, ions, nitrogen-vacancy centres in diamond, superconducting circuits, {\it etc.}). The configuration of these energy levels does not matter: the hybrid Rabi-STIRAP pulses can be applied to the ladder, $\Lambda$, or $V$ configurations. Here we present an analysis of the parameters needed to run the hybrid STIRAP protocol for a particular kind of superconducting circuit, the transmon.

\subsection{Superconducting circuits realizing qutrits}

Superconducting circuits that realize qubits and qutrits can be made with present microfabrication technology. These are artificial atoms governed by a Hamiltonian that comprise a Josephson part and potential energy terms (inductive or capacitive).  The Josephson energy appears whenever two superfluids or superconductors are connected via a weak link: it appears not only in metallic superconductors \cite{Josephson} where it was originally investigated, but also in superfluid $^{3}$He \cite{Josephson3He} and $^{4}$He \cite{Josephson4He}, in Bose-Einstein atomic condensates \cite{ RevModPhys.73.307} as well as in quantum degenerate Fermi gases with interatomic interactions \cite{PhysRevA.66.041603, PhysRevLett.105.225301}. The capacitive (charge) energy results from the junction capacitance itself or from additional capacitors realized on-purpose on the chip. Depending on the value of these capacitors, phenomena such as Coulomb blockade can appear. Such charging effects have found a variety of applications, for example Coulomb thermometry \cite{ PhysRevLett.73.2903, Meschke2016}, sensitive charge detection by single-electron transistors \cite{RevModPhys.64.849, Schoelkopf1998, Paraoanu2003}, on-chip coolers \cite{ PhysRevB.90.201407}, subgap thermometry \cite{ PhysRevB.76.172505}, and thermal machines \cite{ PhysRevLett.115.260602}. The interplay between the charging energy and the Josephson energy in nanoelectronics has been studied since the late ‘80s \cite{ PhysRevLett.63.1307}, leading to the observation of effects such as the resonant tunneling of Cooper pairs \cite{PhysRevLett.73.1541,PhysRevB.76.172505}.
The transmon \cite{transmon_PRA2007}, which is the focus of the present work, combines in a clever way the Josephson and charging energy to realize a slightly anharmonic multilevel system which at the same time is immune to spurious charge fluctuations due to nonequilibrium electrons \cite{Riste13}.
Thus one can identify in this system three or more levels in the ladder configuration, which are stable and addressable by microwave field.

\subsection{Effective Hamiltonian in a circuit QED setup}

\begin{figure}[tb]
\centering
\includegraphics[width=15cm]{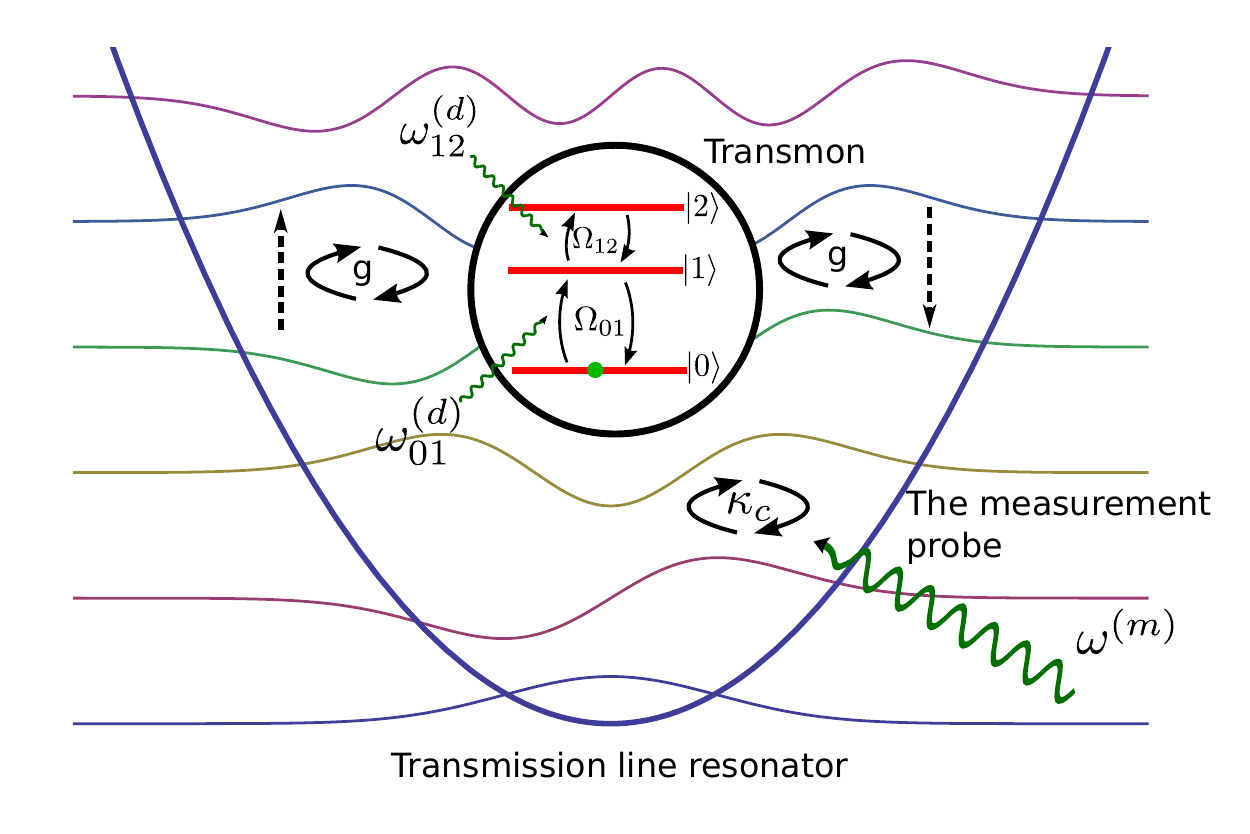}
\caption{In a circuit QED architecture, the three-level system (for example a transmon) is dispersively coupled to the modes of a cavity, which is used as a read-out. In this configuration, one can perform full quantum tomography of the three-level system.}
\label{circuitQED}
\end{figure}

The proposed experiment can be realized in a typical circuit QED architecture - with the transmon embedded (with generic coupling strength $g$) in a cavity with decay rate $\kappa_{c}$ \cite{Schuster05}. To neglect the effects of the environment, we assume that the coherence time of the transmon is larger than the duration of the protocol (usually of the order of tens to hundreds of nanoseconds), a situation that has already been achieved in experiments.  Typically the three-level system is far-off resonant with respect to the cavity, and in this dispersive regime the cavity can be used as a readout device (interrogated with a probe measurement of frequency $\omega^{(m)}$), see Fig. \ref{circuitQED}.
Indeed, due to the coupling between the transmon and the resonator cavity, the resonant frequency of the later $\omega_r$ changes depending on the state of the three-level system. This effect is used for quantum tomography in the following way. A probe pulse at fixed frequency $\omega^{(m)}$ and fixed amplitude is sent to the resonator, and the reflected or transmitted signal is downconverted and recorded in time domain. The shape of the recorded trace $r(\tau)$ depends on the qubit state, and this allows to map the state of the transmon into a specific trace. Once the calibration traces are recorded $r_j(\tau), j = {0,1,2}$, where $j$ corresponds to preparing the transmon in either of $|0\rangle$, $|1\rangle$ or $|2\rangle$ states, it is possible to decompose the trace $r(\tau)$ of any superposition  of the above three basis states with some weights $p_j$, where $p_j$ is the probability to find the system in the state $|j\rangle$. Thus, given a recorded trace $r(\tau)$, we can identify the occupation probabilities $p_j$.

When the transmon is irradiated with two fields resonant to the first and second transition, the Hamiltonian reads

\begin{equation}
\label{eq:jc}
\begin{aligned}
H =& \hbar \omega_0 |0\rangle \langle 0| + \hbar \omega_1 |1\rangle \langle 1| + \hbar \omega_2 |2\rangle \langle 2| +
\frac{\hbar \Omega_{01}}{2} \left(\sigma_{01}^{\dag} e^{-i\omega_{01}^{(d)}t} + h.c. \right) +
\frac{\hbar \Omega_{12}}{2} \left(\sigma_{12}^{\dag} e^{-i\omega_{12}^{(d)}t} + h.c. \right) \\
 & + \frac{\hbar \Omega_{(01)\rightarrow (12)}}{2} \left(\sigma_{12}^{\dag} e^{-i\omega_{01}^{(d)}t} + h.c. \right) +
\frac{\hbar \Omega_{(12)\rightarrow (01)}}{2} \left(\sigma_{01}^{\dag} e^{-i\omega_{12}^{(d)}t} + h.c. \right) ,
\end{aligned}
\end{equation}
where $\Omega_{(01)\rightarrow (12)}$ and $\Omega_{(12)\rightarrow (01)}$ are the cross-couplings of the 0-1 pulse into the 1-2 transition and of the 1-2 pulse into the 0-1 transition respectively. The Pauli annihilation and creation operators for the 0-1 and 1-2 transitions are defined as $\sigma_{01} = |0\rangle \langle 1|$ and $\sigma_{12} = |1\rangle \langle 2|$. For the particular case of the transmon in the harmonic approximation, the cross-couplings are given by $\Omega_{(01)\rightarrow (12)} = \sqrt{2} \Omega_{01}$ and $\Omega_{(12)\rightarrow (01)} = (1/\sqrt{2}) \Omega_{12}$.

Next, we take the drives resonant with the corresponding transitions ($\omega_{01}^{(d)} = \omega_1 - \omega_0 =\omega_{01}$ and
$\omega_{12}^{(d)} = \omega_2 - \omega_1 =\omega_{12}$), then
move to a rotating frame and perform the rotating wave approximation. Now, if
$|\omega_{01}^{(d)} - (\omega_2 - \omega_1)|,|\omega_{12}^{(d)} - (\omega_1 - \omega_0)| \gg \Omega_{(01)\rightarrow (12)},\Omega_{(12)\rightarrow (01)}$, then we can neglect the cross-coupling terms. In this case, we are left precisely with our starting Hamiltonian Eq. (\ref{eq:threelevel}).

\subsection{Hybrid pulses under realistic experimental conditions}

We discuss here the nonidealities caused by the driving. With resonant driving fields, we have $\omega_{01}^{(d)} = \omega_{01},\ \omega_{12}^{(d)} = \omega_{12}$, then $\abs{\omega_{01}^{(d)} - (\omega_2 - \omega_1)} = \abs{\omega_{01} - \omega_{12}}$ and $\abs{\omega_{12}^{(d)} - (\omega_1 - \omega_0)} = \abs{\omega_{12} - \omega_{01}} = \abs{\omega_{01} - \omega_{12}}$. Therefore both expressions refer to the anharmonicity of a three-level system. For the transmon this gives $\abs{\omega_{01}^{(d)} - (\omega_2 - \omega_1)} = \abs{\omega_{12}^{(d)} - (\omega_1 - \omega_0)} = E_c/\hbar$ in the first order approximation, which is typically around $2\pi\cdot 300\ \textrm{MHz}$.

Next, we estimate the cross-coupling strengths for a two-field driving due to the imperfect cancellation of the fast rotating terms under the rotating wave approximation, as in Eq. (\ref{eq:jc}) of the previous subsection.
We take the expressions $\int_{-\infty}^{\infty}|\Omega_{01}(t)|\ dt = \int_{-\infty}^{\infty}|\Omega_{12}(t)|\ dt \simeq 10\pi$,
and extract the maximum driving amplitudes $\Omega_{01}^{(0)} = \Omega_{12}^{(0)}\simeq  5\sqrt{2\pi}/\sigma$. For $\sigma = 50\ \textrm{ns}$ this gives $\Omega_{01}^{(0)}=\Omega_{12}^{(0)}\simeq 2\pi\cdot 40\ \textrm{MHz}$.
With this the cross-couplings in the harmonic approximation will be $\Omega_{(01)\rightarrow(12)} = \sqrt{2}\Omega_{01}^{(0)} \simeq 2\pi\cdot 57\ \textrm{MHz}$, and $\Omega_{(12)\rightarrow(01)} = (1/\sqrt{2})\Omega_{12}^{(0)}\simeq2\pi\cdot 28\ \textrm{MHz}$,
 and the above condition takes the form $300 \gg 57,\ 28$. Thus, for the parameters used here it is possible to neglect these terms, however, due to the low anharmonicity of the transmon this approximation becomes worse if we need to use higher powers.

Using the Hamiltonian Eq. (\ref{eq:jc}) we can study in detail the situation when the cross-driving terms are not negligible. This is relevant for systems with low anharmonicity (such as the transmon) or for systems with small coherence times (in which case one has to increase the Rabi and STIRAP strengths in order to achieve a fast enough operation). Surprisingly, we find that the phase relations obtained in the ideal case (without cross-coupings) in Section 2 are not altered much: the relative phase $\mathrm{Arg}(AB^*)$ as plotted in Fig. 4 left panel remains unchanged $\mathrm{Arg} (AB^*) = - \pi /2 + \phi_{01}$, while the relative phase $\mathrm{Arg} (AC^{*})$ from Fig. 3 right panel only acquires a constant shift $\mathrm{Arg}(AC^{*}) = \mathrm{const.} + \mathrm{Arg}(\beta ) - \mathrm{Arg}(\alpha ) + 2 \phi_{01} + \phi_{12}$. This demonstrates again the robustness of our method.




\section{Conclusions}

We have shown that hybrid pulses consisting of standard nonadiabatic Rabi pulses followed by Raman sequences can be used to create arbitrary superpositions between the three states of a qutrit. The concepts developed here apply to any physical realization of the qutrit. We have demonstrated that the population of the target state $|2\rangle$ of STIRAP is controlled exclusively by the initial Rabi pulse, while the population of the initial $|0\rangle$ and intermediate $|1\rangle$ state depends also on the width of the STIRAP pulse, in agreement with previous experimental results. To achieve full quantum control, we have analyzed the phase differences of the amplitudes of the qutrit states. We have found that, for a given pulse area, any phase difference can be realized by varying the phases of the two STIRAP pulses in a relatively simple way: the phase difference added by the STIRAP pulses with phases $\phi_{01}$ and $\phi_{12}$ turn out to be $\phi_{01}$ (between the initial state and the intermediate state) and $2\phi_{01} + \phi_{12}$ (between the initial and final states). This holds only in the adiabatic regime - when the separation between pulses becomes either too long or too short, these simple relations are no longer satisfied. We also have shown that if a fourth state (ancillary state) is available, we can use it to find nondestructively the phase differences. Finally, we have identified the conditions under which these protocols can be realized experimentally in circuit QED setups.

\vspace{6pt}


\acknowledgments{This work used the cryogenic facilities of the Low Temperature Laboratory at Aalto University. We acknowledge financial support from the Center for Quantum
Engineering at Aalto University (project QMET), V\"aisal\"a Foundation, and the the Academy of Finland (project 263457 and project 250280 - Center of Excellence ``Low Temperature Quantum Phenomena and Devices'' ).}

\appendix

\section{General three-level superposition as initial state for STIRAP}
\label{appendix}

Our focus in this work has been on a single nonadiabatic pulse that prepares a superposition of the states $|0\rangle$ and $|1\rangle $ that serves as the initial state for STIRAP. However, in general we can consider an arbitrary three-level superposition of the type
\begin{equation}
|\psi (t_\tau ) = \alpha |0\rangle + \beta |1\rangle + \gamma |2\rangle \notag
\end{equation}
as the input state for the STIRAP: this can be realized at some time $t_{\tau}$ for example by using two Rabi pulses (one acting on the $0 - 1$ transition and the other on the $1 - 2$ transition). In this case, we have instead of Eq. (\ref{fracSTIRAP}) the following expression for the state after a fractional STIRAP sequence
\begin{eqnarray}
|\psi (t)\rangle = \left[ \alpha \cos \Theta (t) - i \beta e^{i \phi_{01}}\sin \Theta (t) \sin \left(\frac{1}{2} \int_{t_{\tau}}^{t}dt' \Omega (t' )\right)
+ \gamma \sin \Theta (t) \cos \left(\frac{1}{2}\int_{t_{\tau}}^{t}dt' \Omega (t' )\right) e^{i \phi_{01} + i \phi_{12}}
\right] |0\rangle &\notag\\
+ \left[\beta \cos \left(\frac{1}{2}\int_{t_{\tau}}^{t}dt' \Omega (t' )\right) - i \gamma \sin \left(\frac{1}{2} \int_{t_{\tau}}^{t}dt' \Omega (t' )\right) e^{i \phi_{12}}\right]
|1\rangle &\notag\\
 -\left[ \alpha e^{-i \phi_{01}-i \phi_{12}}\sin \Theta (t)  + i \beta \cos \Theta (t) \sin \left(\frac{1}{2} \int_{t_{\tau}}^{t}dt' \Omega (t' ) \right)e^{-i \phi_{12}} + \gamma \cos \Theta (t) \cos \left(\frac{1}{2}\int_{t_{\tau}}^{t}dt' \Omega (t' )\right)
\right] |2\rangle &. \notag
\end{eqnarray}
For a full STIRAP we then obtain
\begin{eqnarray}
\label{eq:psi_gen}
|\psi (\infty )\rangle &=& \left[- i \beta e^{i \phi_{01}}\sin \left(\frac{1}{2} \int_{t_{\tau}}^{\infty}dt' \Omega (t' ) \right)
+ \gamma \cos \left(\frac{1}{2}\int_{t_{\tau}}^{t}dt' \Omega (t' )\right) e^{i \phi_{01} + i \phi_{12}}
\right]|0\rangle \notag\\
& & + \left[\beta \cos \left(\frac{1}{2}\int_{t_{\tau}}^{\infty}dt' \Omega (t' )\right)- i \gamma \sin \left(\frac{1}{2} \int_{t_{\tau}}^{t}dt' \Omega (t' )\right) e^{i \phi_{12}}\right]|1\rangle \notag \\
& & - \alpha e^{-i (\phi_{01}+\phi_{12})}|2\rangle . \notag
\end{eqnarray}
The result is somewhat unexpected: we see that the population of state $|2\rangle$ does not depend on the initial amplitude $\gamma$. If we attempt to ``cheat'' by first transferring population on $|2\rangle$, then applying STIRAP, we will not achieve a higher final population.





\bibliographystyle{apsrev4-1}
\bibliography{ref_hybridSTIRAP}


\end{document}